\documentclass{WileyMSP-template}
\usepackage{amsthm,amsmath}
\usepackage[utf8]{inputenc}
\usepackage{graphicx}
\usepackage{soul, color, xcolor}
\begin{document}

\pagestyle{fancy}
\rhead{\includegraphics[width=2.5cm]{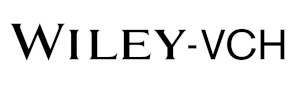}}

\title{Weak force sensing based on optical parametric amplification in a cavity optomechanical system coupled in series with two oscillators.}

\maketitle


\author{Zheng Liu}
\author{Yu-qiang Liu}
\author{Yi-jia Yang}
\author{Chang-shui Yu*}
\begin{affiliations}
	
	Zheng Liu\\
	Address\\Dalian University of Technology
	Email Address:liuzheng@mail.dlut.edu.cn
	
	Yu-qiang Liu\\
	Address\\Dalian University of Technology
	Email Address:lyqmyt@163.com
	
   Yi-jia Yang\\
   Address\\Dalian University of Technology
   Email Address:yangyijia@mail.dlut.edu.cn
	
   Chang-shui Yu\\
   Address\\Dalian University of Technology
   Email Address:ycs@dlut.edu.cn
\end{affiliations}



\begin{abstract}

In the realm weak force sensing, an important issue is to suppress fundamental noise (quantum noise and thermal noise), as they limit the accuracy of force measurement. In this paper, we investigate a weak force sensing scheme that combines a degenerate optical parametric amplifier (OPA) and an auxiliary mechanical oscillator into a cavity optomechanical system to reduce quantum noise. We demonstrate that the noise reduction of two coupled oscillators depends on their norm mode splitting. and provide a classic analogy and quantum perspective for further clarification. Besides, the noise reduction mechanism of OPA is to reduce the fluctuation of photon number and enhance the squeezing of the cavity field. We propose a specific design aimed at enhancing the joint effect of both, beyond what can be achieved using OPA alone or two series coupled oscillators. This scheme provides a new perspective for deeper understanding of cavity field squeezing and auxiliary oscillator in force sensing.

\end{abstract}


\section{Introduction}
\label{I} 
Recent advancements in the theory and technology of quantum precision measurement have significantly improved the stability and accuracy of physical quantity measurement \cite{RevModPhys.89.035002,RevModPhys.82.1155,10.3389/fphy.2022.1035435,PhysRevLett.128.011802,PhysRevLett.128.220504,PhysRevLett.129.070502,Zhang2024}. For instance, the stability of femtosecond time transmission, which could redefine the second in the International System of Units (SI), has been achieved \cite{shen2022free}. Moreover, high-sensitivity gravitational wave signal measurements have been realized using squeezed light \cite{PhysRevLett.129.121103}. These achievements underscore the superior capabilities of quantum sensing.

The cavity optomechanical system, which integrates optical detection and an oscillator as a probe \cite{PhysRevLett.127.113601}, presents a natural platform for quantum precision measurement. Numerous realms have illustrated its advantages, such as weak force sensing \cite{wsf}, acceleration sensing \cite{add}, weak field detection schemes \cite{zhang}, and biosensors \cite{8528407,s20061569,2020Optomechanical}, These applications have considerable practical implications for the fabrication of quantum sensors. Among them, in the field of weak force sensing, the basic idea is to couple the optical cavity with a mechanical oscillator through the radiation pressure. When the external force acts on the mechanical oscillator, it will change its displacement and modulate the frequency of cavity field. Therefore, the displacement information of the mechanical oscillator can be inferred from the optical quadrature output spectrum, indirectly reflecting the weak force signal. Even in low-temperature environments where thermal noise can be ignored, however, quantum noise, originating from Heisenberg's uncertainty principle, will continue to limit the sensitivity of quantum sensing. In cavity optomechanical sensing realm, The typical sources of quantum noise are shot noise and backaction noise \cite{li2021cavity}. Therefore, many weak force sensing schemes aim to reduce quantum noise. These include introducing squeezed light \cite{PhysRevLett.115.243603,aasi2013enhanced,Zhao2019,Zhang:24,10.1063/5.0208107}, coherent quantum noise cancellation (CQNC) \cite{Motazedifard_2016}, quantum non-demolition sensing schemes \cite{PhysRevA.98.043804}.

Recent cavity optomechanics research has proposed numerous schemes to enhance the performance of the cavity optomechanical system by utilizing nonlinear optical devices such as the Kerr medium \cite{PhysRevA.101.023841}, optical parametric amplifier (OPA) \cite{PhysRevA.102.023503}, and four-wave mixing \cite{PhysRevA.97.033806}. The integration of these nonlinear devices with the optomechanical system can lead to intriguing phenomena like mechanical oscillator squeezing \cite{opazhenziqiangyasuo}, optomechanical entanglement enhancement \cite{opaguanglijiuchan}, and improved ground state cooling effect for mechanical oscillators \cite{opajitailengque,Liao:22}. In particular, the strong nonlinearity induced by OPA can effectively enhance the sensitive detection precision \cite{PhysRevA.95.023844}.

In this paper, we introduce an OPA and an auxiliary oscillator into the standard cavity optomechanical system, combined with a homodyne detection device, to explore the performance of weak force sensing. As a third degree of freedom, the auxiliary oscillator can provide norm mode splitting, reduce quantum noise, and simultaneously shift the frequency of optimal noise suppression. This mechanism can be compared with the motion of the center of mass and the relative motion of classical coupled harmonic oscillators. In addition, By utilizing the pump gain of OPA, the optical cavity field can be amplified to further reduce quantum noise. Compared to schemes using only an auxiliary oscillator or OPA, our scheme can more effectively improve the performance of weak force sensing when under proper parameter conditions.

The structure of the remainder of this paper is as follows: Section II introduces the physical mode for weak force sensing and presents the system's Hamiltonian. Section III provides the quantum Langevin equation for the system and derives the corresponding noise spectral density. Section IV conducts a numerical analysis of the noise spectral density , offering potential physical explanations to elucidate the analytical results. Finally, Section V presents the conclusions, discusses potential experimental implementations, and explores future prospects.
\section*{2 The force sening model and the Hamiltonian}
\label{sec2}
\begin{figure}[!htbp]
	\centering\includegraphics[width=0.7\columnwidth]{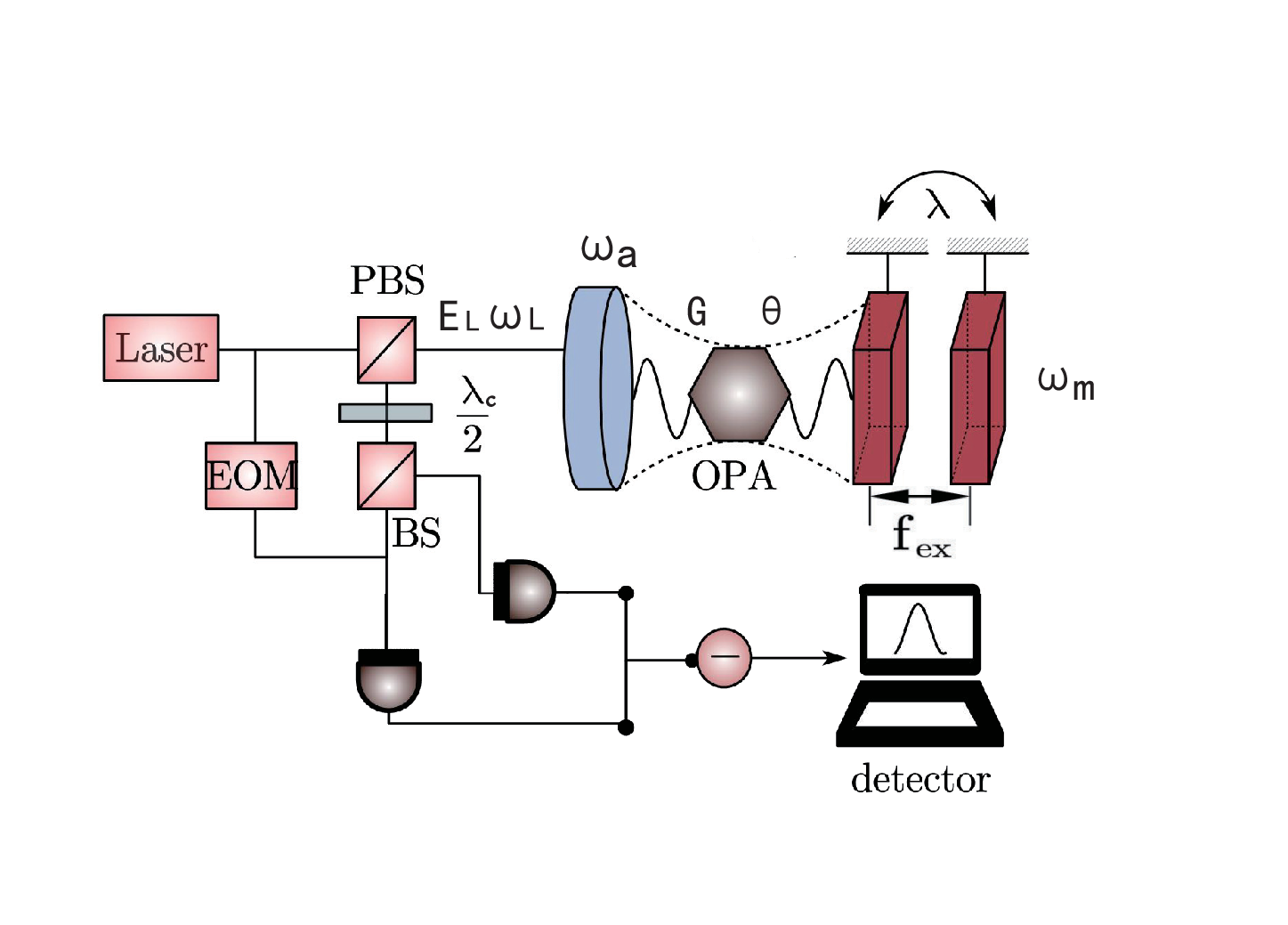}
	\caption{The schematic diagram of the weak force sensing model. This system consists of an optomechanical cavity with an optical parametric amplifier (OPA) and an auxiliary oscillator, where the two mechanical oscillators are coupled in series. A weak detected external force $f_{ex}$ is imposed on the coupled mechanical oscillators. The output of the cavity field through the homodyne detection device can detect the weak force signal. The phase of local oscillating light can be modulated by an electro-optical modulator (EOM).}
	\label{FIG1}
\end{figure}

We consider a cavity optomechanical system including an OPA and two coupled mechanical oscillators with frequencies $\omega _{m1}, \omega _{m2}$ and effective mass $m_{1}, m_{2}$, shown in Fig. \ref{FIG1}. The cavity field with frequency $\omega _{a}$ decays with the rate $\kappa $ and the dissipation rates of the two mechanical oscillators are denoted by $\gamma _{m_{1}}, \gamma _{m_{2}}$. We use an overcoupled ($\kappa $$\approx $$\kappa _{ex}$ ) cavity in this model with $\kappa _{ex} $ denoting the external dissipation rate. A pump field with amplitude $E_{L}$ and frequency $\omega _{L}$ drives the cavity, and a pumping field of frequency $2\omega _{L}$ drives the OPA with the nonlinear gain $G$ and the pump phase $\theta$. When the free spectral range (FSR) of the cavity is much larger than the mechanical frequency $\omega _{m_{i}}$, the single-mode approximation can be safely used \cite{cklaw}, which implies the interaction will not cause light scattering to other modes. Thus, in the frame rotating at the driving field frequency $\omega _{L}$, the total Hamiltonian of the force sensing system reads
\begin{align}
	\hat{H}^{\prime }& =\hbar \Delta _{a}\hat{a}^{\dagger }\hat{a}+\frac{\hbar
		\omega _{m1}}{2}(\hat{X}_{1}^{2}+\hat{P}_{1}^{2})+\frac{\hbar \omega _{m2}}{2%
	}(\hat{X}_{2}^{2}+\hat{P}_{2}^{2})  \notag \\
	& +\hbar g_{0}\hat{a}^{\dagger }\hat{a}\hat{X}_{1}+\hbar \lambda \hat{X}_{1}%
	\hat{X}_{2}+i\hbar E_{L}(\hat{a}^{\dagger }-\hat{a})  \notag \\
	& +i\hbar G(e^{i\theta }\hat{a}^{\dagger 2}-e^{-i\theta }\hat{a}^{2}).
	\label{(1)}
\end{align}
The first three terms of the Hamiltonian denote the free Hamiltonian of the cavity field and two oscillators, respectively, where $\hbar $ is the reduced Planck constant, $\Delta _{a}=\omega _{a}-\omega _{L}$ is the detuning between the cavity field and the driving field, $a(a^{\dagger })$ is the bosonic annihilation (creation) operator of the cavity field, and $\hat{X}_{i}=\frac{\hat{x}_{i}}{X_{i0}},\hat{P}_{i}=\frac{\hat{p}_{i}}{P_{i0}}$ are dimensionless position and momentum operators of the mechanical oscillators with $X_{i0}=\sqrt{\frac{\hbar }{m\omega _{mi}}},P_{i0}=\sqrt{m\omega _{mi}\hbar }$ representing the zero-point fluctuations of the position and momentum of the oscillators. The last four terms of the Hamiltonian describe the coupling between the cavity field and the first oscillator, the interaction of the two oscillators, and the coherent driving field on the cavity mode and the OPA, respectively, where $g_{0}=\omega _{a}X_{10}/L$ is the radiation pressure coupling coefficient of a single photon with the cavity length $L$, $\lambda $ is the coupling coefficient of the two oscillators, and ${|E_{L}|=\sqrt{\frac{\kappa P_{L}}{\hbar \omega _{L}}}}$ with $P_{L}$ representing the input power of the coherent driving field.
\section*{3 The dynamics of the force sensing system}

\label{sec3}

In this section, we solve the dynamics of weak force sensing systems with the aim of providing analytical expressions for quantities that characterize sensing performance. We use the Heisenberg-Langevin equation to solve the dynamics
of our system. According to the Hamiltonian (\ref{(1)}), one can directly
write Heisenberg-Langevin equations \cite{PhysRevA.105.033507} as
\begin{align}
	&\dot { \hat { a } } = - ( i \Delta _ { a } + \frac {\kappa}
	{ 2 } ) \hat { a } - i g _ { 0 } \hat { a } \hat X _ { 1 } + E _ { L } + 2 G
	e ^ { i \theta } \hat { a } ^ { \dagger } + \sqrt { \kappa } \hat { a } _ {
		i n } , \\&\dot { \hat {X _ { i }} } = \omega_ { m i } \hat { { P } _ { i }
	}, (i=1, 2) \\& \dot { \hat { P } } _ { 1 } = - \omega _ { m _ { 1 } } \hat
	{ X } _ { 1 } - g _ { 0 } \hat { a } ^ { \dagger } \hat { a } - \gamma _ { m
		_ { 1 } } \hat { P } _ { 1 } - \lambda\hat X _ { 2 } + \sqrt { 2 \gamma _ {
			m _ { 1 } } }f _ { in{ 1 } } ,\\ &\dot { \hat { P } }_ { 2 } = - \omega _ {
		m { 2 } } \hat { X } _ { 2 } - \lambda \hat { X } _ { 1 } - \gamma _ { m _ {
			2 } } \hat { P } _ { 2 } + \sqrt { 2 \gamma _ { m _ { 2 } } } f _ { in  { 2
	} } ,  \label{HL}
\end{align}
where $\hat{a} _ { in }$ is the input noise operator for the cavity field
with the correlation function \cite{haosanlengque} $\langle \hat{a} _ { 
in } ( t ) \hat{a} _ { in } ^ { \dagger } ( t ^ { \prime } )
\rangle=\delta ( t - t ^ { \prime } )$, $\quad f_{ini}=f_ {ex}+f_{thi} (i=1,2) $ is
the input force on the first oscillator, $f _ { t h i } = \frac { F _ { t hi } } {
\sqrt { 2 \hbar m \gamma _ { m i } \omega _ { m i } } }$ is the random force
of dimensionless Brownian motion, which, under the high environment temperature limit,
meets $\langle f_{\text {thi} } (t) f_{\text {thi} } (t^{\prime})\rangle
\approx \bar {n}_{thi} \delta (t-t ^{\prime})$ with $\bar{n} _ { thi } = [
\exp ( \hbar \omega _ { mi } / k _ { B } T ) - 1 ] ^ { - 1 } \approx k _ { B
} T/ \hbar \omega _ { mi }$ the phonon occupation number in thermal
equilibrium, and $f _ { e x } = \frac { F _ { e x t } } { \sqrt { 2 \hbar m
\gamma _ { m i } \omega _ { m i } } }$ is the dimensionless external weak force to be
measured. Due to the fact that the equations is a set nonlinear equations, it is difficult to solve directly. Fortunately, in the case of strong driving and weak optomechanical coupling, standard linearization can be used, and we can substitute the operators by the steady-state average and the first-order fluctuation \cite{Kippenberg:07}, i.e., $\hat { a } = \alpha+ \delta \hat {a},\hat { X } _ { i } = \bar { X } _ { i } + \delta \hat { X } _ { i },\hat { P } _ { i } = \bar { P } _ { i } + \delta \hat { P } _ { i }$. Thus
the above Heisenberg-Langevin equation can be rewritten as
\begin{align}  \label{2}
	&\delta\dot { \hat { a } } = - ( i \Delta^ {\prime}_a + \frac {\kappa} {2})
	\delta\hat { a } - i \alpha g _ { 0 } \delta\hat { X } _ { 1 }+\sqrt {
		\kappa } \delta\hat { a } _ { i n } + 2 G e ^ { i \theta } \delta\hat { a }
	^ {\dagger},  \notag \\
	&\delta\dot {\hat X }_ { i }= \omega_ { m i } \delta\hat {{P}_{i}} , i = 1,2,
	\notag \\
	&\delta\dot { \hat {{P}_{1}}} = - \omega _ { m _ { 1 } } \delta\hat { X }_{
		1 } - g_{ 0 }(\alpha^\ast \delta\hat { a }+\alpha\delta\hat{ a }^{\dagger})
	-\lambda \delta\hat{ X }_{ 2 } - \gamma_{ m_{1}}\delta\hat{P}_{1}  \notag \\
	& + \sqrt {2\gamma_{m_{1}}} f_{ i n 1},  \notag \\
	&\delta\dot {\hat {P_{ 2 } }} = - \omega_{ m_{2} } \delta\hat{X}_{ 2} -
	\lambda \delta\hat{X}_{ 1 } - \gamma_{ m_{2} } \delta\hat{ P }_{ 2 } +
	\sqrt {2 \gamma_{ m_{ 2 }} } f_{in_{ 2 } },
\end{align}
where $\Delta^ { \prime }_a = \Delta _ { a } + g _ { 0 } {\bar X_ { 1 } }$
is the effective detuning after linearization treatment, and the
steady-state average, is given by
\begin{align}  \label{4}
	&\alpha = \frac { ( - i \Delta ^ { \prime }_a +\kappa/2 + 2 G e ^ { i \theta
		} ) E_{L} } {( \Delta ^ { \prime }_a ) ^ { 2 } + \kappa ^ { 2 } / 4 - 4 G ^
		{ 2 } }=|\alpha|e^{i\phi}  \notag \\
	&\bar{ X } _ { 1 } = \frac { - g _ { 0 } | \alpha | ^ { 2 } \omega _ { m 2 }
	} { \omega _ { m 1 } \omega _ { m 2 } - \lambda ^ { 2 } },\bar{ P } _ { 1 }
	= 0,  \notag \\
	&\bar { X } _ { 2 } = \frac { \lambda g _ { 0 } | \alpha | ^ { 2 }  } { \omega _ { m 1 } \omega _ { m 2 } - \lambda ^ { 2 }  }, \bar{ P } _ { 2 }=0,
\end{align}
With the phase of the cavity field $\phi = \arctan \left( \frac { 4 G \sin \theta - 2 \Delta _a ^ { \prime } } { 4 G \cos \theta + \kappa} \right)$, which shows an implicit connection with the OPA pump phase $\theta$. Considering the quadrature components of the cavity mode in the sense of $\hat{x}_{a} = (\hat{a}^{\dagger} + \hat{a})/\sqrt{2}$, $\hat{p}_{a} = (\hat{a} - \hat{a}^{\dagger})/\sqrt{2}i$, $\hat{x}_{a}^{in} = (\hat{a}_{in}^{\dagger} + \hat{a}_{in})/\sqrt{2}$, $\hat{p}_{a}^{in} = (\hat{a}_{in} - \hat{a}_{in}^{\dagger})/\sqrt{2}i$. Under the Routh-Hurwitz stability condition \cite{wendingxing}, $G< 0.25\kappa$ needs to be met. The linearized Heisenberg-Langevin equation (\ref{2}) can be written in compact matrix form as
\begin{equation}
	\dot{m}(t)=Bm(t)+c(t),
\end{equation}
where $ m(t)=[\delta \hat{X}_{1}(t),\delta \hat{P}_{1},\delta \hat{X}_{2},\delta \hat{P}_{2},\delta \hat{x}_{a},\delta \hat{p}_{a}]^{T} $, $ c(t)=[0,\sqrt{2\gamma _{m1}}f_{in1},0,\sqrt{2\gamma _{m2}}f_{in2},\sqrt{\kappa }x_{a}^{{in}},\sqrt{\kappa }p_{a}^{{in}}]^{T} $, and 
\begin{equation}
	B=\left(
	\begin{array}{cccccc}
		{0} & {\omega _{m1}} & {0} & {0} & {0} & {0} \\
		{-\omega _{m1}} & {-{\gamma _{m1}}} & {-\lambda } & {0} & {-g\cos\phi } & { -g\sin\phi } \\
		{0} & {0} & {0} & {\omega _{m2}} & {0} & {0} \\
		{-\lambda } & {0} & {-\omega _{m2}} & {-\gamma _{m2}} & {0} & {0} \\
		{g\sin\phi } & {0} & {0} & {0} & {l_{1}} & {l_{2}} \\
		{-g\cos\phi } & {0} & {0} & {0} & {l_{2}^{\prime }} & {l_{1}^{\prime }}%
	\end{array}%
	\right),
\end{equation}
 with $ l_{1}=(2G\cos \theta -\kappa /2) $, $ l_{1}^{\prime }=(-2G\cos \theta -\kappa /2) $, $ l_{2}=(2G\sin \theta +\Delta _{a}^{\prime }) $, $ l_{2}^{\prime }=(2G\sin \theta -\Delta _{a}^{\prime }) $, and $ g=\sqrt{2}\alpha g_{0} $ is the linearized single photon radiation pressure coupling coefficient. In the latter sections, we will omit $ \delta $ and the `hat' of the fluctuation operators for simplicity.
 \begin{figure}[!htbp]
 	\centering \includegraphics[width=0.8\columnwidth]{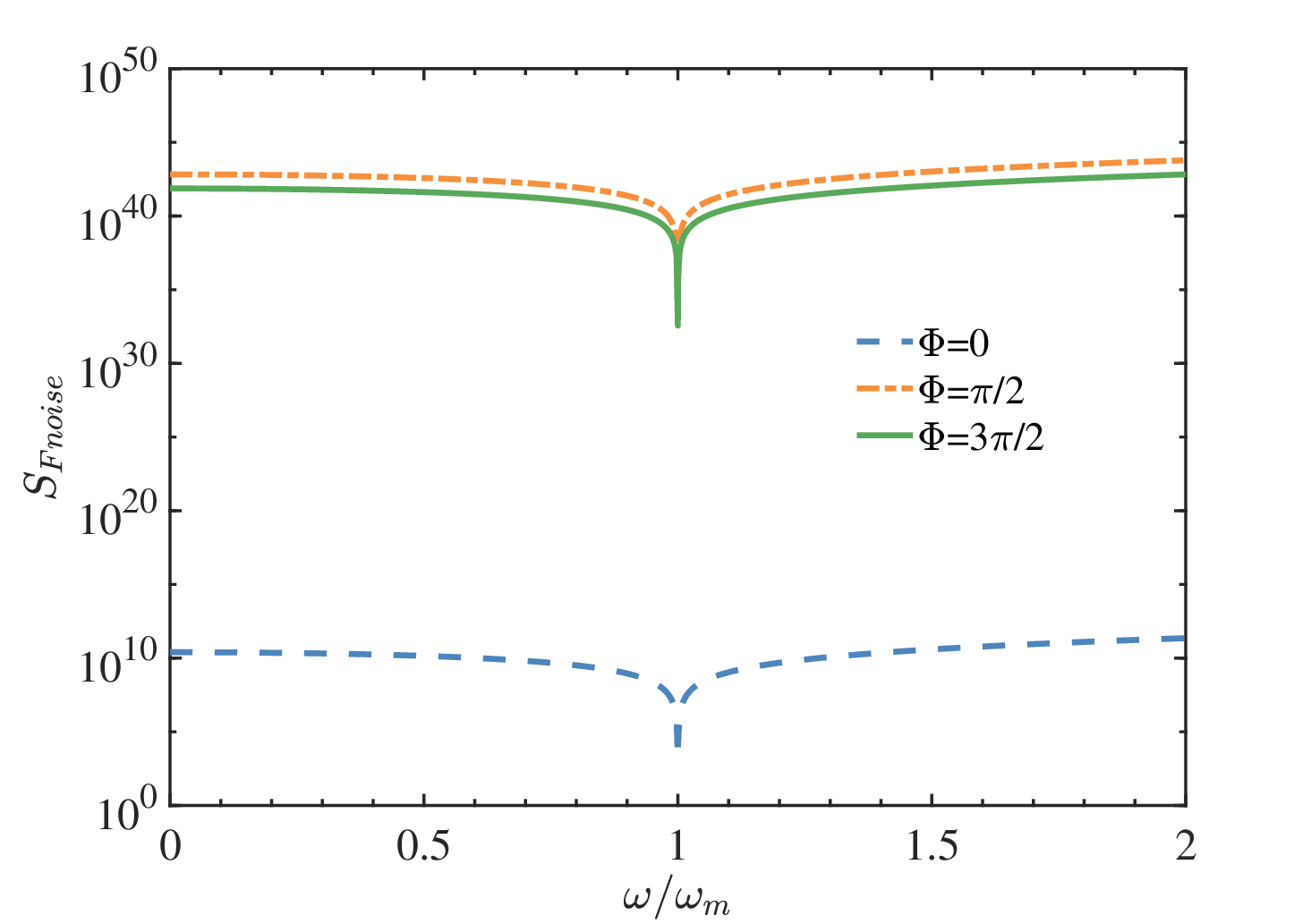}
 	\caption{The dimensionless noise power spectral density $S_{Fnoise}$ versus $\protect\omega $/$\protect\omega _{m}$ for different phases $\Phi $. Other parameters are set as $G$ = 0, $\protect\kappa =10^{2}\protect\omega _{m}$, $\protect\omega _{m1}=\protect\omega _{m2}=2\pi\times10^6Hz$, $T=300K$. }
 	\label{FIG2}
 \end{figure}

In order to better analyze the spectrum of signals and noise, we turn the Heisenberg-Langevin equation to the frequency domain based on the Fourier transform $ {\ f } ( \omega ) = \int _ { - \infty } ^ { \infty } d t \frac { e ^ { i \omega t } } { \sqrt { 2 \pi } } f ( t ) $, which yields 
	\begin{align}
	& { x}_{a}=\chi _{a}^{\prime }(k_{1}{ p}_{a}+k_{2}{f}_{in{2}}+k_{3}{ f}%
	_{in1}+\sqrt{\kappa }x_{a}^{in}), \\
	& { p}_{a}=p_{a}^{\prime }(k_{4}{ x}_{a}+k_{5}{ f}_{in{2}}+k_{6}{ f}%
	_{in1}+\sqrt{\kappa }p_{a}^{in}),
\end{align}%
with
\begin{align}
	&\chi _{a}^{\prime }=1/(-i\omega +\kappa /2-2G\cos \theta +g^{2}\chi
	_{m1}^{\prime }\sin \phi \cos \phi ),  \notag \\
	&p_{a}^{\prime }=1/(-i\omega +\kappa /2+2G\cos \theta -g^{2}\chi
	_{m1}^{\prime }\sin \phi \cos \phi ),  \notag \\
	&k_{1}=2G\sin \theta +\Delta _{a}^{\prime }-g^{2}\sin ^{2}\phi \chi
	_{m1}^{\prime },  \notag \\
	&k_{2}=-\chi _{m1}^{\prime }g\sin \phi \lambda \chi _{m2}\sqrt{2\gamma _{m2}%
	},  \notag \\
	&k_{3}=g\sin \phi \chi _{m1}^{\prime }\sqrt{2\gamma _{m1}},  \notag \\
	&k_{4}=2G\sin \theta -\Delta _{a}^{\prime }+g^{2}\cos ^{2}\phi \chi
	_{m1}^{\prime },  \notag \\
	&k_{5}=\chi _{m1}^{\prime }g\cos \phi \lambda \chi _{m2}\sqrt{2\gamma _{m2}}%
	,  \notag \\
	&k_{6}=-g\cos \phi \chi _{m1}^{\prime }\sqrt{2\gamma _{m1}},
\end{align}
where $\chi _ { m 1 } ^ { \prime } = \frac { \chi _ { m 1 } } { 1 - \lambda
	^ { 2 } \chi_ { m 1 } \chi _ { m 2 } } $ is the
equivalent {susceptibility} of the coupled harmonic oscillator and $%
\chi_{mi}=\frac { \omega _ { m i } } { \omega _ { m _ { i } } ^ { 2 } -
	\omega^ { 2 } - i \omega \gamma_ { m i } }, (i=1,2)$ is the 
susceptibility of two mechanical oscillators.
The external detected force signal is reflected in the output spectrum of the cavity field. To
detect the external force, we intend to choose the optimal generalized
quadrature of the output cavity field.  So we can use the standard input-output relationship $ { x } _ { a } ^ { {out } }= \sqrt { \kappa }  { x } _ { a } -
{ x } _ { a } ^ { i n },  { p } _ { a } ^ { o u t } = \sqrt {\kappa }  { p } _ { a } - { {p } _ { a }}^{ i n }$, 
one can get the quadratures of the output field as
\begin{align}
		{x}_{a}^{out}& =\chi _{aeff}[k_{1}^{\prime }{\ f}_{in2}+k_{2}^{\prime }{\ f}%
		_{in{1}}+\chi _{a}^{\prime }k_{1}p_{a}^{\prime }\sqrt{\kappa }{\ p}_{a}^{in}
		\\
		& +(\sqrt{\kappa }\chi_{a}^{\prime }-\chi _{aeff}^{-1}){x}_{a}^{in}],  \notag \\
		{p}_{a}^{out}& =\chi _{aeff}[k_{3}^{\prime }{\ f}_{in2}+k_{4}^{\prime }{\ f}%
		_{in{1}}+\chi _{a}^{\prime }k_{4}p_{a}^{\prime }\sqrt{\kappa }{x}_{a}^{in}
		\\
		& +(\sqrt{\kappa }p_{a}^{\prime }-\chi _{aeff}^{-1}){p}_{a}^{in}],  \notag
	\end{align}%
	where
	\begin{align}
		& \chi _{aeff}=\sqrt{\kappa }/(1-\chi _{a}^{\prime }k_{1}p_{a}^{\prime
		}k_{4}),  \notag \\
		& k_{1}^{\prime }=\chi _{a}^{\prime }k_{1}p_{a}^{\prime }k_{5}+\chi
		_{a}^{\prime }k_{2},  \notag \\
		& k_{2}^{\prime }=\chi _{a}^{\prime }k_{1}p_{a}^{\prime }k_{6}+\chi
		_{a}^{\prime }k_{3}.  \notag \\
		& k_{3}^{\prime }=\chi _{a}^{\prime }k_{2}p_{a}^{\prime }k_{4}+p_{a}^{\prime
		}k_{5},  \notag \\
		& k_{4}^{\prime }=\chi _{a}^{\prime }k_{3}p_{a}^{\prime }k_{4}+p_{a}^{\prime
		}k_{6}.
\end{align}
The generalized quadrature written as the linear combination of the
amplitude and phase of the output field can be given by \cite%
{vitali,lingchatance}
\begin{align}
		{p}_{a,\Phi }^{out}& =\cos \Phi {p}_{a}^{out}-\sin \Phi {x}_{a}^{out}  \notag
		\\
		& =\chi _{aeff}[(k_{3}^{\prime }\cos \Phi -k_{1}^{\prime }\sin \Phi ){\ f}%
		_{in2}  \notag \\
		& +(k_{4}^{\prime }\cos \Phi -k_{2}^{\prime }\sin \Phi ){\ f}_{in1}  \notag
		\\
		& +(\chi _{a}^{\prime }k_{4}p_{a}^{\prime }\sqrt{\kappa }\cos \Phi -(\sqrt{%
			\kappa }\chi _{a}^{\prime }-\chi _{aeff}^{-1})\sin \Phi ){x}_{a}^{in}  \notag
		\\
		& +((\sqrt{\kappa }p_{a}^{\prime }-\chi _{aeff}^{-1})\cos \Phi -\chi
		_{a}^{\prime }k_{1}p_{a}^{\prime }\sqrt{\kappa }\sin \Phi ){p}_{a}^{in}],
\end{align}
where $\Phi $ is the phase of the local oscillating light, which can be modulated by EOM. The input force ${\ f}_{in1}$ can be divided into the external measured force and the total noise force ${f}_{in1}={\ f}_{ex}(\omega )+{\ F}_{noise,\Phi }(\omega )$. Correspondingly, the dimensionless total noise force operator can be obtained as
	\begin{align}
		& {F}_{noise,\Phi }=[\frac{k_{4}^{\prime }\cos \Phi -k_{2}^{\prime }\sin
			\Phi }{k_{4}^{\prime }\cos \Phi -k_{2}^{\prime }\sin \Phi+k_{3}^{\prime }\cos \Phi -k_{1}^{\prime }\sin
			\Phi }{\ f}%
		_{th1} +\frac{k_{3}^{\prime }\cos \Phi -k_{1}^{\prime }\sin
			\Phi }{k_{4}^{\prime }\cos \Phi -k_{2}^{\prime }\sin \Phi+k_{3}^{\prime }\cos \Phi -k_{1}^{\prime }\sin
			\Phi }{\ f}_{th2} \notag \\
		& +\frac{\chi _{a}^{\prime }k_{4}p_{a}^{\prime }\sqrt{\kappa }\cos \Phi -(%
			\sqrt{\kappa }\chi _{a}^{\prime }-\chi _{aeff}^{-1})\sin \Phi }{k_{4}^{\prime }\cos \Phi -k_{2}^{\prime }\sin \Phi+k_{3}^{\prime }\cos \Phi -k_{1}^{\prime }\sin
			\Phi }{x}_{a}^{in} +\frac{(\sqrt{\kappa }p_{a}^{\prime }-\chi _{aeff}^{-1})\cos \Phi -\chi
			_{a}^{\prime }k_{1}p_{a}^{\prime }\sqrt{\kappa }\sin \Phi }{k_{4}^{\prime }\cos \Phi -k_{2}^{\prime }\sin \Phi+k_{3}^{\prime }\cos \Phi -k_{1}^{\prime }\sin
			\Phi  }{p}_{a}^{in}].
\end{align}
Based on the definition of the symmetrical noise spectral density \cite{PhysRevA.89.053836}
\begin{equation}
	S_{Fnoise}(\omega )\delta (\omega +\omega ^{\prime })=\frac{1}{2}(\langle {\
		F}_{noise}(\omega ){\ F}_{noise}(\omega ^{\prime })\rangle +c.c.)
\end{equation}%
we can utilize the correlation
functions of the quadratures as \cite{zhengjiaoguanlian}

\begin{align}
	& \langle x_{a}^{{\ in}}(\omega )x_{a}^{{\ in}}(\omega ^{\prime })\rangle
	=\langle p_{a}^{{\ in}}(\omega )p_{a}^{{\ in}}(\omega ^{\prime })\rangle =%
	\frac{1}{2}\delta (\omega +\omega ^{\prime }), \\
	& \langle x_{a}^{{\ in}}(\omega )p_{a}^{{\ in}}(\omega ^{\prime })\rangle
	=-\langle p_{a}^{{\ in}}(\omega )x_{a}^{{\ in}}(\omega ^{\prime })\rangle =%
	\frac{i}{2}\delta (\omega +\omega ^{\prime }),
\end{align}%
and the dimensionless noise spectral density is given as
\begin{align}
		& S_{Fnoise,\Phi }=|\frac{k_{4}^{\prime }\cos \Phi -k_{2}^{\prime }\sin
			\Phi  }{k_{4}^{\prime }\cos \Phi
			-k_{2}^{\prime }\sin \Phi +k_{3}^{\prime
			}\cos \Phi -k_{1}^{\prime }\sin \Phi} |^{2}\frac{k_{B}T}{\hbar \omega _{m1}}+|\frac{k_{3}^{\prime
			}\cos \Phi -k_{1}^{\prime }\sin \Phi }{k_{4}^{\prime }\cos \Phi
			-k_{2}^{\prime }\sin \Phi+k_{3}^{\prime
			}\cos \Phi -k_{1}^{\prime }\sin \Phi }|^{2}\frac{k_{B}T}{\hbar \omega _{m2}}  \notag \\
		& +\frac{1}{2}|\frac{\chi _{a}^{\prime} k_{4}p_{a}^{\prime }\sqrt{\kappa }%
			\cos \Phi -(\sqrt{\kappa }\chi _{a}^{\prime }-\chi _{aeff}^{-1})\sin \Phi }{%
			k_{4}^{\prime }\cos \Phi -k_{2}^{\prime }\sin \Phi +k_{3}^{\prime
			}\cos \Phi -k_{1}^{\prime }\sin \Phi}|^{2} +\frac{1}{2}|\frac{(\sqrt{\kappa }p_{a}^{\prime }-\chi _{aeff}^{-1})\cos
			\Phi -\chi _{a}^{\prime }k_{1}p_{a}^{\prime }\sqrt{\kappa }\sin \Phi }{%
			k_{4}^{\prime }\cos \Phi -k_{2}^{\prime }\sin \Phi +k_{3}^{\prime
			}\cos \Phi -k_{1}^{\prime }\sin \Phi}|^{2}.
\end{align}
The first and second terms represent the thermal noise of the two
oscillators, and the third and fourth terms represent the backaction noise and
shot noise of the cavity field, respectively. In addition, the noise spectral
density reflects the knowability of noise when measuring weak force, so it
can also be regarded as the sensitivity of weak force sensing \cite%
{zhengjiaoguanlian}. 

\section*{4 Sensing performance of weak force sensing system}
\label{IV}
In this section, we introduce the sensing performance of our system, including its response to external signals, suppression of additional noise, and signal-to-noise ratio. By analyzing these performances, we can study the influence of various physical parameters on the sensing characteristics of the system. In addition, to simplify the analysis, we selected two identical harmonic oscillators, namely  $\omega _{m1}=\omega _{m2}=\omega _{m}$, $\gamma _{m_{1}}=\gamma _{m_{2}}=\gamma _{m}$.
\begin{figure}[!htbp]
	\centering \includegraphics[width=0.8\columnwidth]{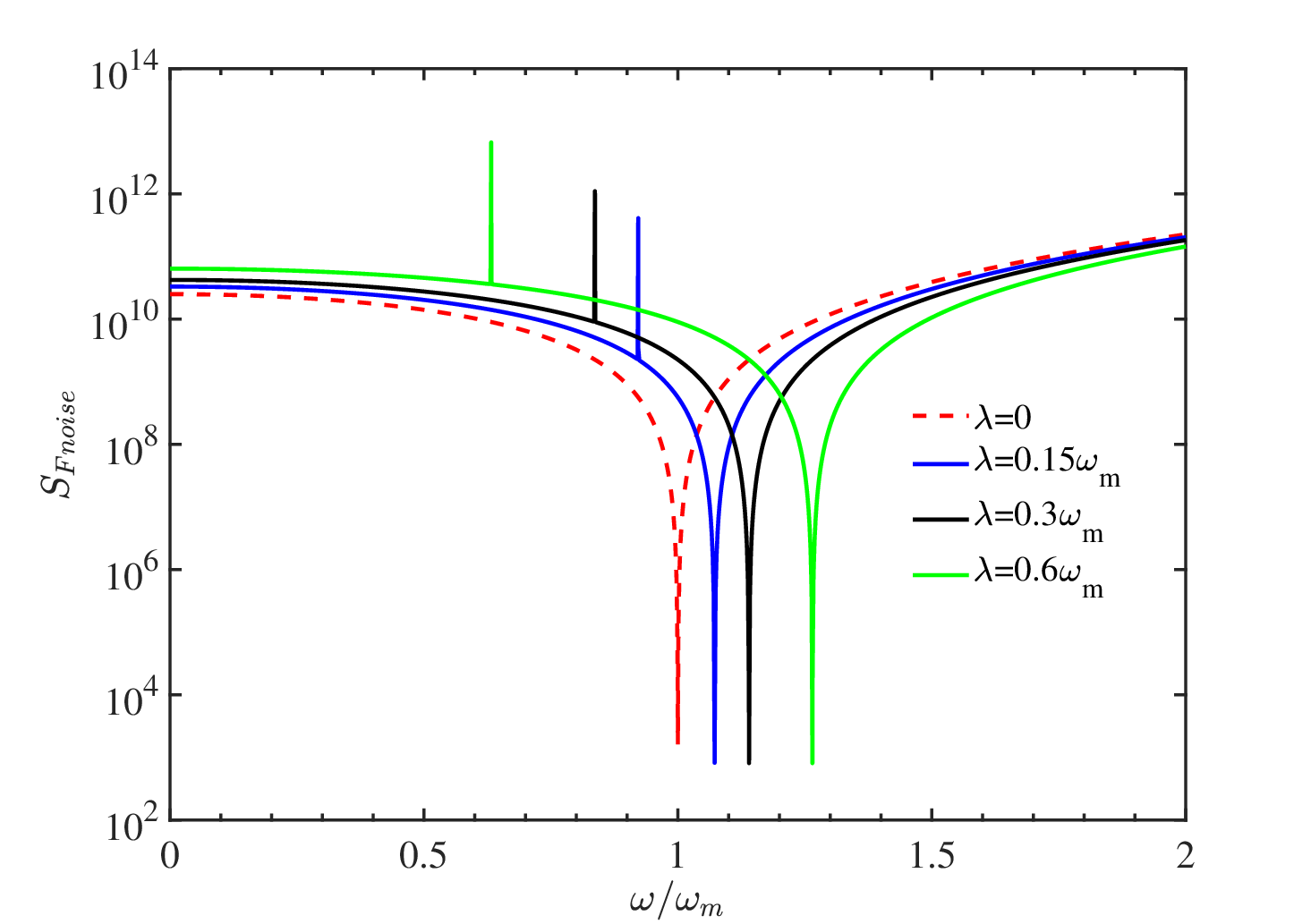}
	\caption{The dimensionless noise power spectral density $S_{Fnoise}$ versus $\protect\omega $/$\protect\omega _{m}$ with low thermal noise, other parameters are set as $G$ = 0, $\protect\kappa =10^{2}\protect\omega_{m}$, $\protect\omega _{m1}=2\pi\times10^6Hz$, $\gamma _{m}=10^{-5}\omega_m$. The ambient temperature is set to $77mK$.}
	\label{FIG3}
\end{figure}
\subsection*{4.1 Coupling with auxiliary oscillator}

In Fig. \ref{FIG2}, we plot the noise power spectral density versus the frequency without OPA and auxiliary oscillator taken into account. One can see that the sensitivity varies with the phase $\Phi $ and the best sensitivity is achieved at $\Phi $ $=0$, since this angle leads to the minimal total noise spectral density. In this sense, we will take the homodyne detection angle $\Phi $ $=0$ in the later discussion. In addition, for simplicity, we set the equivalent detuning $\Delta^ { \prime }_a$ = 0, $\phi$ = 0. Next, we consider the effect of auxiliary oscillator and OPA on noise suppression. According to the experimental parameters, $\omega_m = 2\pi\times1.04\times10^{6}Hz$, $Q_m = 6.2\times10^5$, the temperature of the environment can reach $T = 77 \rm mK$ \cite{Zhang_2017}, which means that we can safely set the thermal noise of the weak force sensing system to a very small value allowed by the experiment, so that we can pay more attention to the suppression of quantum noises. In Fig. \ref{FIG3}, compared with the case without auxiliary oscillator (i.e., $\lambda =0$), we find that the auxiliary oscillator induces two norm modes in the noise spectrum density, especially when $\lambda =0.15\omega _{m}$; $\omega \approx 1.07\omega _{m}$. The effect of noise suppression is greater than that without auxiliary oscillator for $\omega =\omega _{m}$.
\begin{figure}[!htbp]
	\centering \includegraphics[width=0.6\columnwidth]{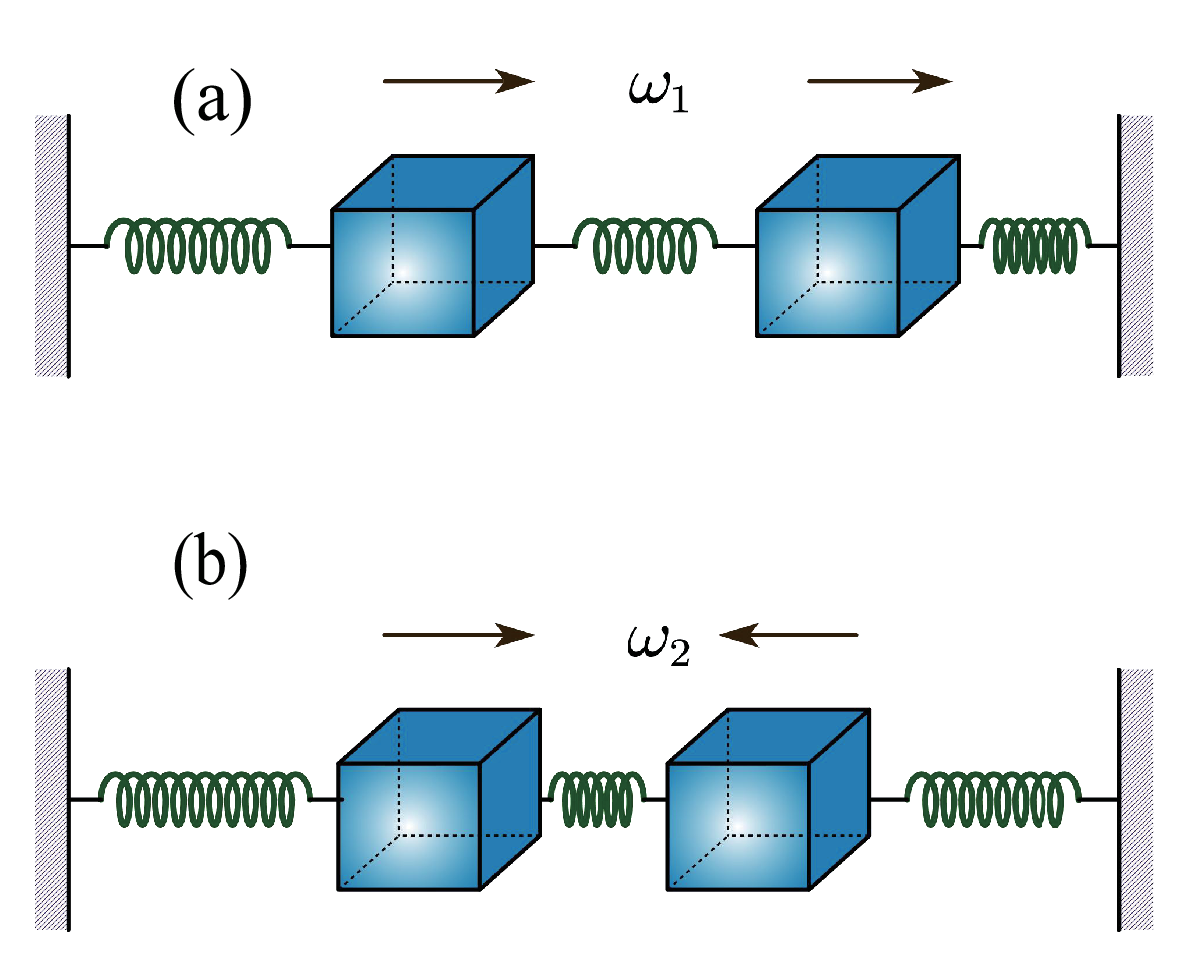}
	\caption{Simplified X-X model diagram of classical one-dimensional harmonic oscillator. (a) Centroid motion. (b) Relative motion.}
	\label{FIG4}
\end{figure}
The existence of two extreme points in the noise spectral density can be effectively explained by  the interaction between two oscillators with frequency $\omega_m$. To explain this phenomenon, we rewrite the Hamiltonian of linearized using the Bosonic creation ($\hat{b}_i^{\dagger}$) and annihilation operators ($\hat{b}_i$), with
\begin{align}
	H_{Li}=\hbar \omega_m ( \hat{b}_{1} ^ {\dagger}\hat{b}_{1} + \hat{b}_{2}^{\dagger}\hat{b}_{2} + 1)\nonumber +\frac {\hbar\lambda} { 2 } (\hat{b}_{1} ^ {\dagger}\hat{b}_{2} ^ {\dagger }+\hat{b}_{ 1 } \hat{b}_{ 2 } +\hat{b}_{1} ^ { \dagger } \hat{b} _ { 2 } +\hat{b}_{1}\hat{b}_{2} ^ {\dagger})+\hbar g(\hat{a}^{\dagger }+\hat{a})(\hat{b}_1+\hat{b}^{\dagger}_1).
\end{align}
It is worth noting that we omit the Hamiltonian of OPA, then we can diagonalize the part of the coupled mechanical oscillator by introducing the following transformation
\begin{align}
	\hat{b}_ { 1 }& = \frac { 1 } { 2 \sqrt { 2 } } ( A_ { 1 } \hat{c} _ { 1 } + A _ { 2 }\hat{c} _ { 2 } + A_ { 1 } ^ { - } \hat{c} _ { 1 } ^ { \dagger } + A _ { 2 } ^ { - } \hat{c} _ { 2 } ^ { \dagger } ) ,\nonumber \\  \hat{b} _ { 2 }& = \frac { 1 } { 2 \sqrt { 2 } } ( A _ { 1 }\hat{c} _ { 1 } - A _ { 2 }\hat{c} _ { 2 } + A_ { 1 } ^ { - } \hat{c} _ { 1 } ^ { \dagger } - A _ { 2 } ^ { - } \hat{c} _ { 2 } ^ { \dagger } ) , \nonumber \\  \hat{b} _ { 1 } ^ { \dagger }& = \frac { 1 } { 2 \sqrt { 2 } } ( A _ { 1 } ^ { - } \hat{c} _ { 1 } + A _ { 2 } ^ { - } \hat{c} _ { 2 } + A _ { 1 }\hat{c} _ { 1 } ^ { \dagger } + A _ { 2 } \hat{c} _ { 2 } ^ { \dagger } ) ,\nonumber  \\ \hat{b}_ { 2 } ^ { \dagger }& = \frac { 1 } { 2 \sqrt { 2 } } ( A _ { 1 } ^ { - } \hat{c} _ { 1 } - A _ { 2 } ^ { - }\hat{c} _ { 2 } + A _ { 1 }\hat{c} _ { 1 } ^ { \dagger } - A _ { 2 } \hat{c} _ { 2 } ^ { \dagger } ) .  
\end{align}
The transformation coefficients are as follows
\begin{equation}
	\begin{aligned}
		A _ { i } = \sqrt { \frac { \omega_m } { \omega _ { i } } } + \sqrt { \frac { \omega _ { i } } { \omega _m} } ,  \quad A _ { i } ^ { - } = \sqrt { \frac { \omega_m } { \omega _ { i } } } - \sqrt { \frac { \omega _ { i } } { \omega_m } },
	\end{aligned}
\end{equation}
where 
$\quad \omega _ { 1 } ^ { 2 } = \omega ^ { 2 }_m +  \lambda \omega_m, \quad \omega _ { 2 } ^ { 2 } = \omega ^ { 2 }_m-\lambda \omega_m. $ Ultimately, Simplified Hamiltonian is
\begin{equation}\label{eq25} 
	\hat{H} ^ { \prime }_{Li} = \hbar \omega _ { 1 } ( \hat{c} _ { 1 } ^ { \dagger } \hat{c} _ { 1 } + \frac { 1 } { 2 } ) + \hbar \omega _ { 2 } (\hat{c} _ { 2 } ^ { \dagger } \hat{c} _ { 2 } + \frac { 1 } { 2 } )+\hbar g \sqrt{\frac{\omega_m}{2}}(\hat{a}^{\dagger }+\hat{a})\left[\frac{\hat{c}_1+\hat{c}_1^{\dagger}}{\sqrt{\omega_1}}+\frac{\hat{c}_2+\hat{c}_2^{\dagger}}{\sqrt{\omega_2}}\right]. 
\end{equation} 
This interaction results in the formation of two normal modes, characterized by frequencies $\omega_{1}^{2} = \omega_{m}^{2} - \lambda \omega_{m}$ and $\omega_{2}^{2} = \omega_{m}^{2} + \lambda \omega_{m}$.  $\omega_{1}$ and $\omega_{2}$ correspond precisely to the locations of the Extreme points observed in the noise power spectrum (Fig. \ref{FIG3}). Notably, the noise power spectral density is significantly reduced at the higher frequency normal mode ($\omega_{2}$). This phenomenon is analogous to the centroid and relative vibrations observed in two coupled classical harmonic oscillators, as depicted in Fig. \ref{FIG4}. In such systems, the overall vibration is a superposition of two normal oscillations, where $\omega_{1}$ represents the centroid motion and $\omega_{2}$ the relative motion.  Relative motion in mechanical oscillators, when coupled with a cavity field, can simulate an equivalent single harmonic oscillator, thereby enhancing response to external forces. Conversely, centroid motion can diminish the noise reduction effect.
\begin{figure}[!htbp]
	\centering \includegraphics[width=0.8\columnwidth]{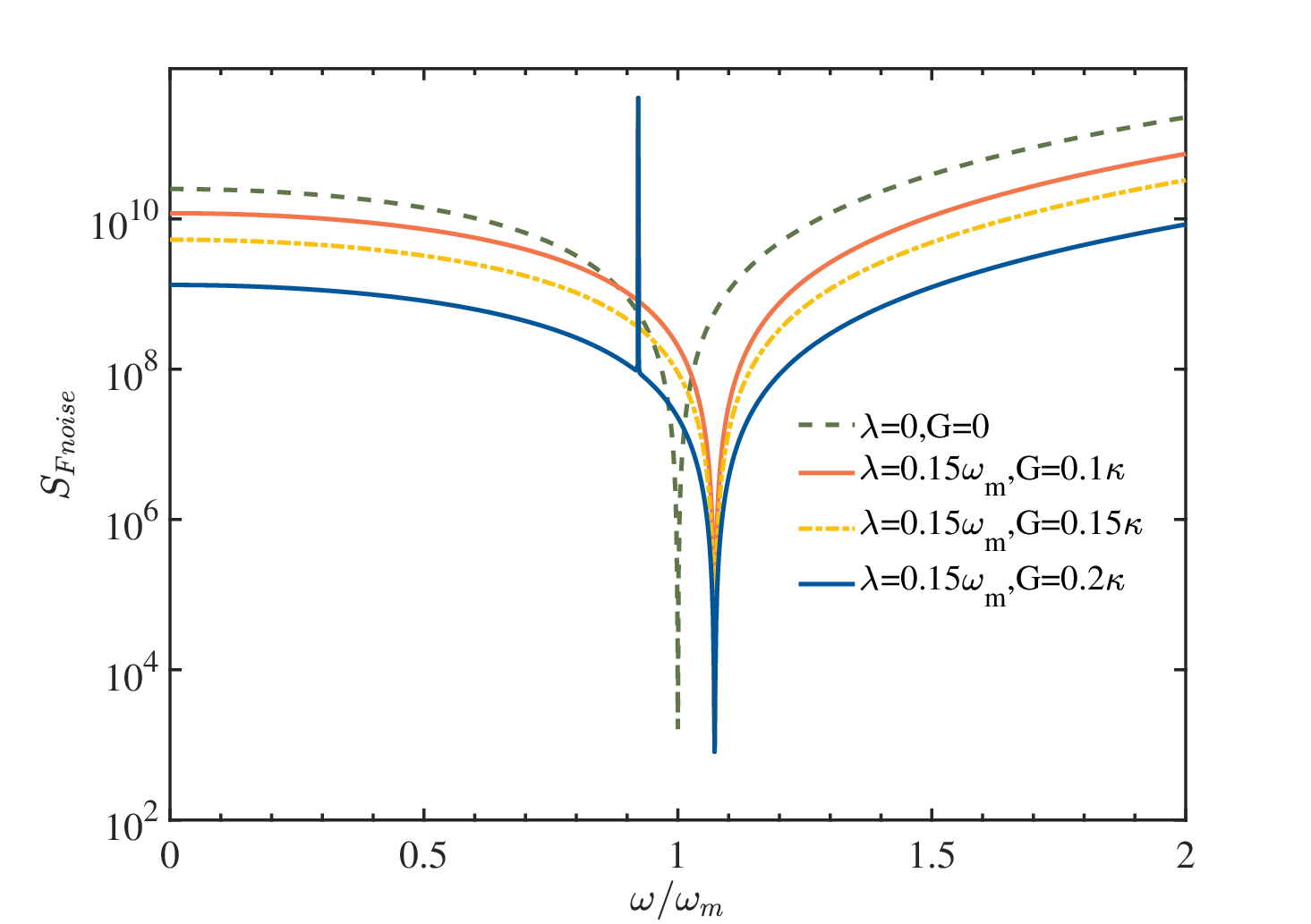}
	\caption{The dimensionless noise power spectral density $S_{Fnoise}$ versus $\omega/\omega_{m}$, for $G$ = 0, 0.15$\kappa$, 0.2$\kappa$, respectively. $\kappa =10^{2}\omega_{m}$, $\omega_{m1}=\omega_{m2}=2\pi\times10^6$Hz, $T=77mK$. }
	\label{FIG5}
\end{figure}
\subsection*{4.2 The role of OPA}
To find out the joint effect of OPA and the auxiliary oscillator, we plot the noise spectral density of the system versus $\omega/\omega_{m}$ under different OPA gain $G$. In Fig. \ref{FIG5}, the OPA can not only improve the sensitivity of the system in the high normal frequency but also strongly increases its bandwidth. Specifically, for $G=0.2 \kappa$, noise across a broad frequency range is greatly suppressed. 
\begin{figure}[!htbp]
	\centering 	\includegraphics[width=15cm,height=10cm]{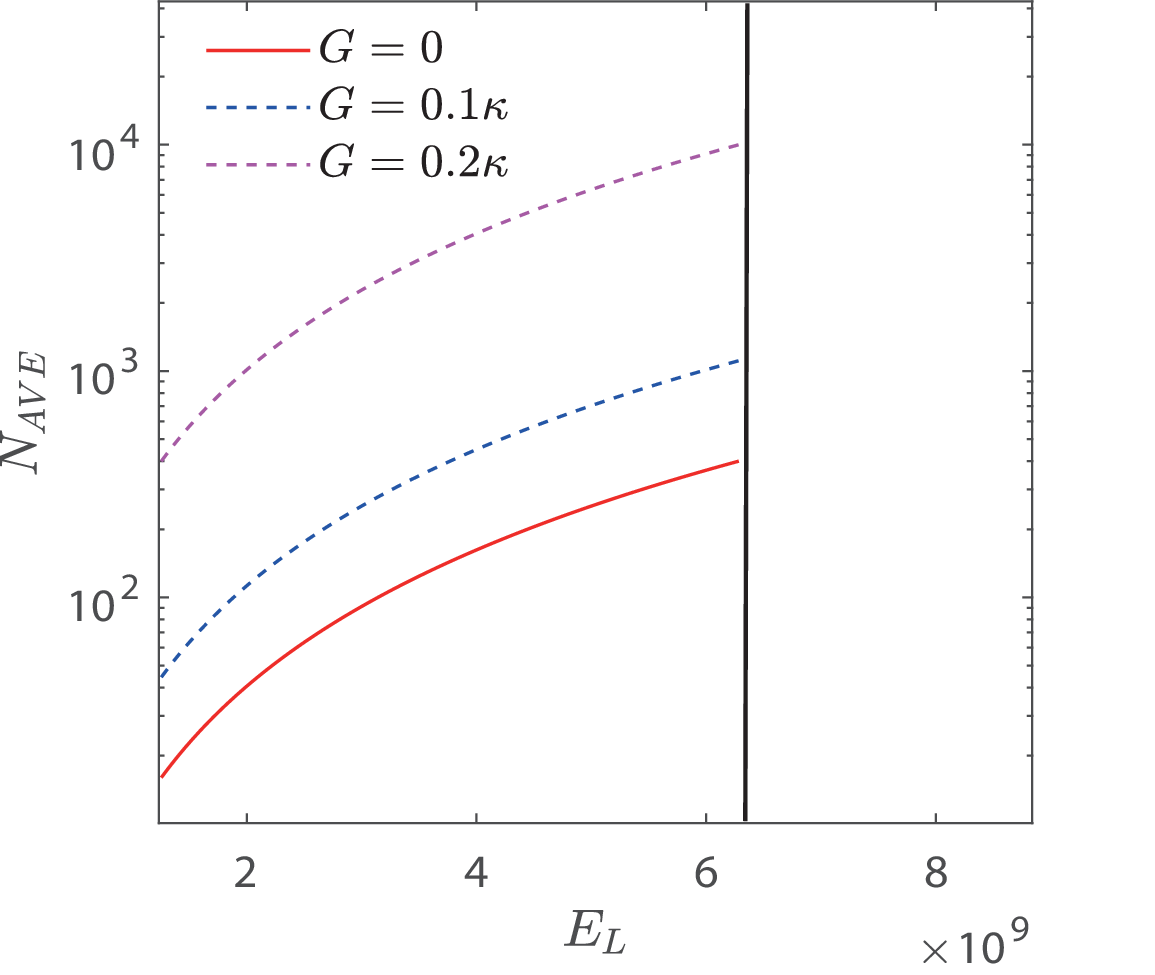}
	\caption{The average photon number in the cavity, denoted as $ N_{\text{AVE}}=|\alpha|^2 $, varies with the drive strength $ E_{L} $ at resonance. For this analysis, we fix $ \theta $ at 0 and explore $ G/\kappa $ ratios of 0, 0.1, and 0.2, maintaining consistency with the parameters used in Figure \ref{FIG5}.}
	\label{FIG6}
\end{figure}
To further understand the noise reduction mechanism of OPA in Fig. %
\ref{FIG6}, we plot the relationship between the average number of photons in
the cavity $N_{AVE}$ and the driving amplitude $E_{L}$. In Fig. \ref{FIG6}, we can see that the average number of photons in the cavity field
increases with the increase of pump gain $G$. As the average number of
photons increases, the fluctuation of the number of photons decreases, which
reduces the shot noise of the system. Meanwhile, with the strong driving
plus the combined action of OPA, the average photon number $N_{AVE}\gg 1$,
so our previous linearized approximation of Heisenberg Langevin equation can
be used. In addition,the OPA exhibits an amplification effect that induces ponderomotive squeezing of the cavity field, thereby elucidating the mechanism of noise suppression. Fig. \ref{FIG7} illustrates the variance of the cavity field's phase quadrature plotted against the pump gain $G$. It is evident that with a fixed coupling strength of the two mechanical oscillators, the quadrature phase component of the cavity field undergoes significant squeezing, diminishing to less than 1/2. However, an increase in the coupling strength between the two mechanical oscillators does not enhance the phase squeezing effect of the cavity field. Hence, it can be understood that the noise reduction mechanism of the auxiliary oscillator may not be due to the squeezing effect, but rather the result of the norm mode splitting effect mentioned earlier.

\begin{figure}[!htbp]
	\centering \includegraphics[width=0.8\columnwidth]{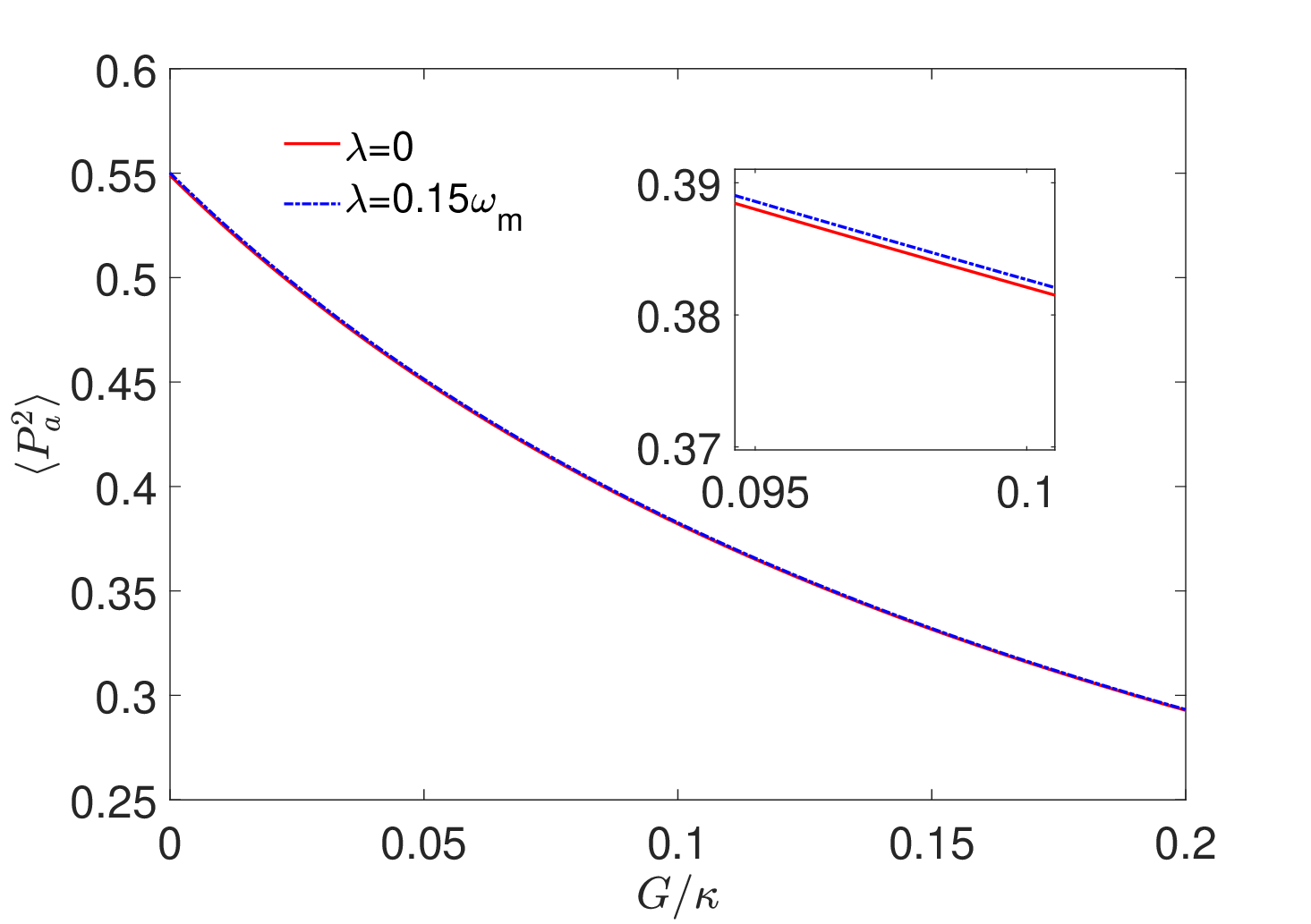}
	\caption{The variance of the phase quadrature component $\langle P^{2}_{a} \rangle$ of the cavity field varies with the pump gain $G$ under different coupling strengths of two oscillators.}
	\label{FIG7}
\end{figure}
\subsection*{4.3 The effect of the optomechanical coupling}
\begin{figure}[!htbp]
	\centering \includegraphics[width=0.8\columnwidth]{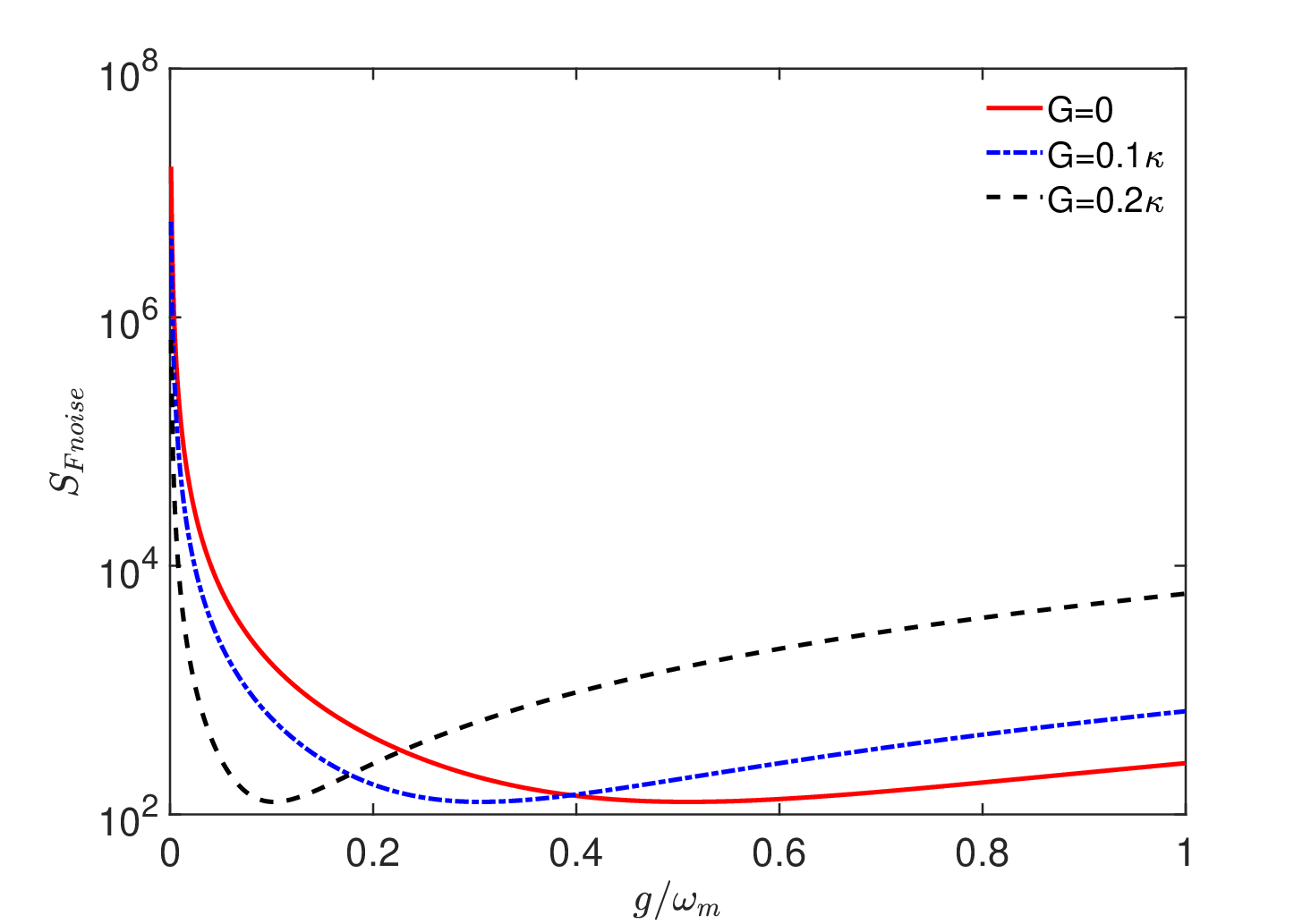}
	\caption{The dimensionless noise spectral density $S_{Fnoise}$ versus $g$. Here, we fix
		the coupling strength $\protect\lambda $ = 0.15$\protect\omega _{m}$, $%
		\protect\omega $ = 1.07$\protect\omega _{m}$, $G$ is taken as 0, 0.1$\protect%
		\kappa $, and 0.2$\protect\kappa $ respectively, and other parameters are $%
		\protect\kappa $=$10^{2}$$\protect\omega _{m}$, $\protect\gamma _{m1}$ = $%
		10^{-5}$$\protect\omega _{m}$.}
	\label{FIG8}
\end{figure}

To intuitively illustrate the influence of optomechanical coupling strength $g$ on noise spectral density, we plot the noise spectrum in Fig.  \ref{FIG8}. One can find that as the coupling strength $g$ increases, the noise reduction effect of OPA gain $G$ becomes the noise enhancement effect. As discussed previously about quantum noise, increasing $g$ will lead to the backaction noise instead of shot noise occupying the dominant position. Therefore, we must carefully adjust the coupling strength $g$ so that the optomechanical interaction in our system is under the weak coupling mechanism ($g_{0}\ll \kappa $ , $\omega _{m}$) \cite{shen2018reconfigurable}, which is feasible under the experiment. 
\subsection*{4.4 The joint effects}
\begin{figure}[!htbp]
	\centering \includegraphics[width=0.8\columnwidth]{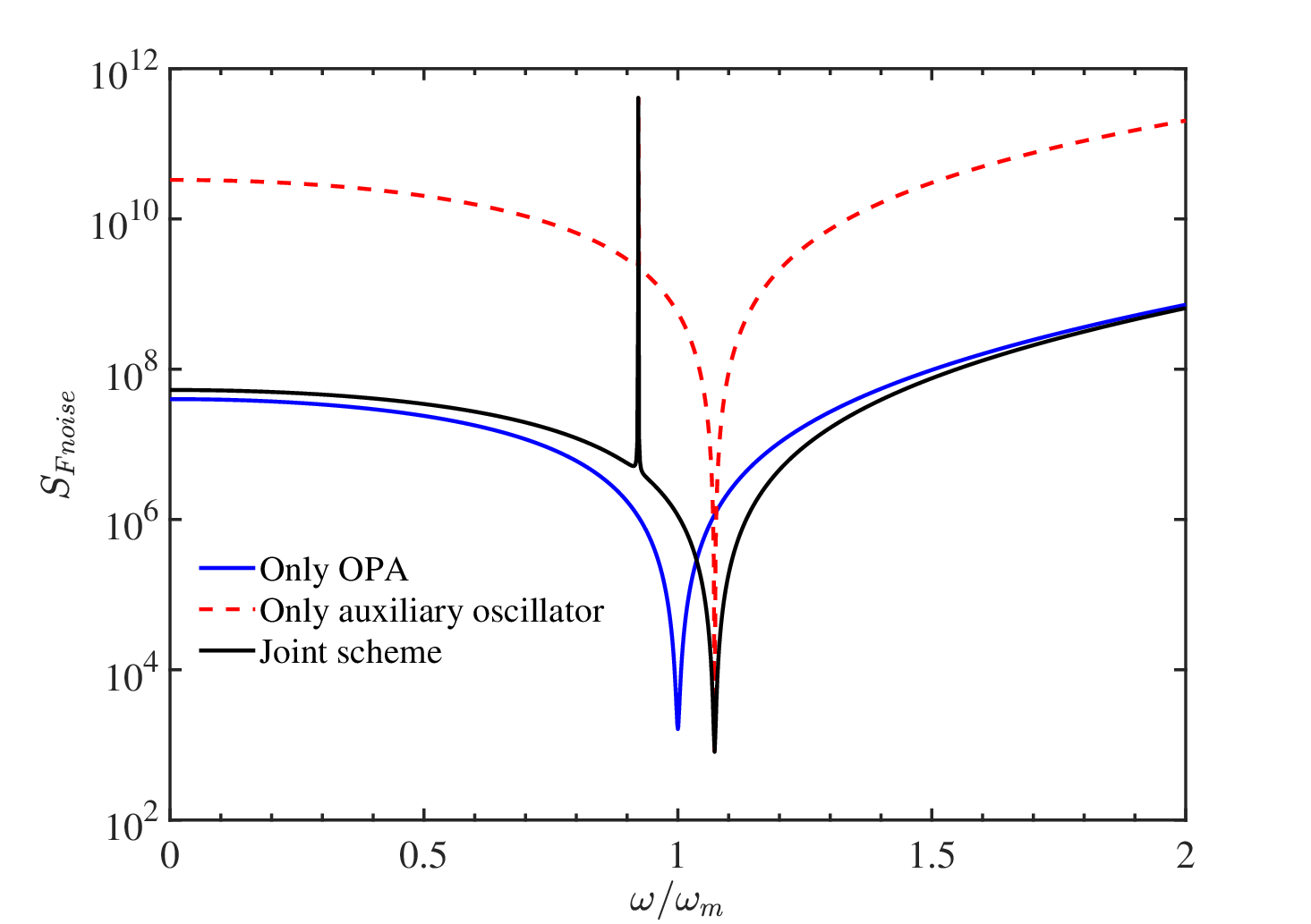}
	\caption{The dimensionless noise power spectral density $S_{Fnoise}$ versus $\protect \omega /$ $\protect\omega _{m}$. The black solid line indicates the scheme of the combination of OPA and auxiliary oscillator ($\lambda=0.15\omega_m$, $G=0.24\kappa$), the blue solid line indicates the scheme of only OPA ($G=0.24\kappa$), and the red dashed line indicates the scheme of only auxiliary oscillator ($\lambda=0.15\omega_m$).}
	\label{FIG9}
\end{figure}

In Fig. \ref{FIG9}, we compare the
joint effect of the OPA combined with the auxiliary oscillator and
their individual effects under the same parameter conditions. We can find that the collective effect can
lead to the stronger optimal suppression of noise at $\omega $ $\approx $
1.07$\omega _{m}$ than any individual effect, and thus has higher sensitivity. 
Besides, the joint scheme have two advantages. On the one hand, it can exceed the sensitivity of individual schemes to achieve higher precision sensitive detection. On the other hand, due to the frequency shift of the optimal noise spectral density caused by the auxiliary oscillator, it has a certain frequency tunability and can adapt to sensing of multiple frequency signals, which has practical significance.

\section*{5 Discussions and conclusions}
\label{V}
In a practical scenario, due to the strong coupling between our mechanical oscillators, it is difficult to achieve strong coupling between mechanical oscillations in experiments. Recently, many theoretical and experimental schemes have been proposed to solve this problem \cite{PhysRevB.86.041402,PhysRevLett.119.053601,PhysRevLett.118.254301,PhysRevLett.117.017701,PhysRevLett.116.103601}, providing experimental feasibility for our scheme. Strong dynamic coupling between two mechanical resonators can be achieved using piezoelectric transducers \cite{PhysRevLett.118.254301,PhysRevLett.117.017701}. The coupling parameters between mechanical oscillators can be controlled by adjusting the gate voltage \cite{PhysRevLett.117.017701}. In addition, two mechanical oscillators can achieve ultra-strong coupling through optical field mediation \cite{PhysRevLett.116.103601,spethmann2016cavity}. Specifically, the coupling strength can reach the magnitude of the intrinsic frequency of the mechanical oscillator ($\lambda\approx\omega_m$) through the indirect action of the optical spring \cite{PhysRevLett.116.103601}. Therefore, under existing experimental conditions, the required coupling strength can be achieved by adding piezoelectric components or optical springs. The effect of OPA can be achieved by devices composed of second-order non-linear crystals \cite{scully1999quantum,boyd2020nonlinear}. In addition, we have noticed a recent optomechanical biosensing experiment that used a disk resonant cavity with an echo wall optical mode and a radial breathing mode. By depositing bacterial particles on a high-frequency disk resonant cavity, frequency splitting caused by bacterial particles can be detected, thereby reflecting the characteristics of bacterial vibration patterns \cite{2020Optomechanical}. And our plan can also be implemented through experiments on this platform. Specifically, if our auxiliary oscillator is replaced by deposited bacteria, similar results will also be produced.

In a word, we studied the weak force sensing in the cavity
optomechanical system with an auxiliary oscillator and OPA. It is shown that
by properly adjusting the coupling strength between the oscillators, the
shot noise can be effectively reduced. However, the minimum noise
spectral density is not concentrated at the resonance frequency of a single
oscillator, but at the higher normal frequency, which results from the norm mode splitting.  When the coupling strength $\lambda $ of the two oscillators is
fixed, additional noise can also be reduced by adjusting the nonlinear gain $%
G$ of OPA. Besides, We did not consider too many issues related to thermal noise because we have assumed the system in a low-temperature environment. Meanwhile, our system does not need the optomechanical strong coupling regime. We believe that this work can be implemented in known two coupled mechanical cantilevers experiment \cite{doi:10.1021/nl902350b}. It is worth noting that in previous works, there have also been some parametric amplification schemes, but these scheme is very different from our scheme, such as linear feedback scheme \cite{PhysRevLett.111.103603}, self-excited oscillation \cite{PhysRevApplied.14.024079}. The former refers to a linear feedback scheme, which has proven ineffective in enhancing the sensitivity of weak force sensing. However, this approach neglects the inherent nonlinearity of the system itself. In contrast, our scheme demonstrates significant nonlinear parametric amplification interactions of optical cavity modes. As for the latter, the self-excited oscillation scheme differs significantly from our approach, despite its similarity to the Optical Parametric Oscillation (OPO) process. We incorporate an Optical Parametric Amplifier (OPA) into the cavity. Generally, in sensing applications, OPA can amplify optical signals' intensity, thereby enhancing sensor sensitivity. By amplifying the light signal, it can elevate the sensor's signal-to-noise ratio, ultimately improving sensing performance. In comparison to the optomechanical self-oscillation method, introducing OPA enhances the optical signal during the signal processing stage, rather than in the driving and control stages of the mechanical resonator. Consequently, the effects of optomechanical self-oscillation methods and the introduction of OPA into optical cavities differ. The former achieves self-oscillation of the mechanical resonator through optomechanical coupling and detects applied force by analyzing disturbances in the self-oscillation trajectory. In contrast, the latter enhances sensor sensitivity by boosting the optical signal. Finally, In the future, we will continue considering coherent
quantum noise cancellation (CQNC) \cite{moller2017quantum}, or use the
entanglement \cite{jiuchan} and squeezing \cite{Momeni_2018} between
multiple oscillators to further enhance the
force sensing sensitivity, and try to carry out quantum sensing research in
other physical systems.

\medskip
\textbf{Supporting Information} \par 
Supporting Information is available from the Wiley Online Library or from the author.

\section*{ Appendix }
In this section, we use the normal mode method to calculate the noise spectral density and compare it with the methods in the main text to enhance the rigor of our method.

Firstly, leveraging the Hamiltonian (\ref{eq25}), we present the linearized Langevin equation for two normal mechanical oscillator modes and the cavity field mode as follows
\begin{align}
	&\dot{x}^{\prime}_{1} = \omega_{1} p^{\prime}_{1},\notag \\
	&\dot{p}^{\prime}_{1} = -\gamma p^{\prime}_{1} - \left(\omega_{m} + \lambda\right) \frac{\omega_{m}}{\omega_{1}} x^{\prime}_{1} - g x_{a} + \sqrt{2 \gamma}\left(2 f_{ex} + f_{th_{1}} + f_{th_{2}}\right), \notag \\
	&\dot{x}^{\prime}_{2} = \omega_{2} p^{\prime}_{2},\notag \\
	&\dot{p}^{\prime}_{2} = -\gamma p^{\prime}_{2} - \left(\omega_{m} - \lambda\right) \frac{\omega_{m}}{\omega_{2}} x_{2} - g x_{a} + \sqrt{2 \gamma}\left(f_{th_{1}} - f_{th_{2}}\right),\notag \\
	&\dot{x}_{a} = \left(2 G - \frac{\kappa}{2}\right) x_{a} + \sqrt{\kappa} x_{a}^{in},\notag \\
	&\dot{p}_{a} = -g\left(\sqrt{\frac{\omega_{m}}{2\omega_{1}}}x^{\prime}_{1} +\sqrt{\frac{\omega_{m}}{2\omega_{2}}} x^{\prime}_{2}\right)- \left(\frac{\kappa}{2} + 2 G\right) p_{a} + \sqrt{\kappa} p_{a}^{in},
\end{align}

where $x^{\prime}_{j} = (\hat{c_{j}}^{\dagger} + \hat{c_{j}})/\sqrt{2}$ and $p^{\prime}_{j} = -i(\hat{c_{j}} - \hat{c_{j}}^{\dagger})/\sqrt{2}$ represent the quadrature components of two normal modes. Additionally, for simplicity, we assume identical oscillators and select optimal parameter conditions as detailed in the main text: $G = 0.24\kappa$, $\Delta^{\prime}_a = 0$, $\lambda = 0.15\omega_{m}$, and $T = 77 \rm mK$. Solving the linearized Langevin equation in the frequency domain yields the analytical expression for the effective force noise operator
\begin{align}
	F^{\prime}_{noise} = A(\omega) x_{a}^{in} + B(\omega) p_{a}^{in} +C(\omega) f_{th_{1}} + D(\omega) f_{th_{2}}
\end{align}
where
\begin{eqnarray}
	A(\omega)& =&\frac{{i g \sqrt{\kappa} \omega_2 \sqrt{\frac{\omega_m}{\omega_1}} \sqrt{\frac{\omega_m}{\omega_2}} (\gamma_m \omega - i \omega^2 + i \lambda \omega_m + i \omega_m^2)}}{{\sqrt{2\gamma_{m}} (4 G - \kappa + 2 i \omega) \omega_m (i \gamma_m \omega + \omega^2 + (\lambda - \omega_m) \omega_m)}} \notag\\&&+ \frac{{i g \sqrt{\kappa} (\gamma \omega \omega_m - i \omega^2 \omega_m - i \lambda \omega_m^2 + i \omega_m^3)}}{{\sqrt{2} \sqrt{\gamma_{m}} (4 G - \kappa + 2 i \omega) \omega_m (i \gamma_m \omega + \omega^2 + (\lambda - \omega_m) \omega_m)}}, \notag\\
	B(\omega)& =& \frac{{(-4 G + \kappa + 2 i \omega) (i \gamma_m \omega + \omega^2 - \omega_m (\lambda + \omega_m))}}{{4 g \sqrt{\kappa\gamma_m} \omega_m}},\notag\\
	C(\omega) &=& \frac{1}{2} \left(1 - \frac{\sqrt{\frac{\omega_m}{\omega_1}} (-i \gamma \omega - \omega^2 + \omega_m (\lambda + \omega_m))}{\sqrt{\frac{\omega_m}{\omega_2}} (i \Gamma \omega + \omega^2 + (\lambda - \omega_m) \omega_m)}\right),	\notag\\
	D(\omega) &=& \frac{1}{2} \left(1 + \frac{\sqrt{\frac{\omega_m}{\omega_1}} (-i \gamma \omega - \omega^2 + \omega_m (\lambda + \omega_m))}{\sqrt{\frac{\omega_m}{\omega_2}} (i \gamma \omega + \omega^2 + (\lambda - \omega_m) \omega_m)}\right).	
\end{eqnarray}

\begin{figure}[!htbp]
	\centering \includegraphics[width=0.8\columnwidth]{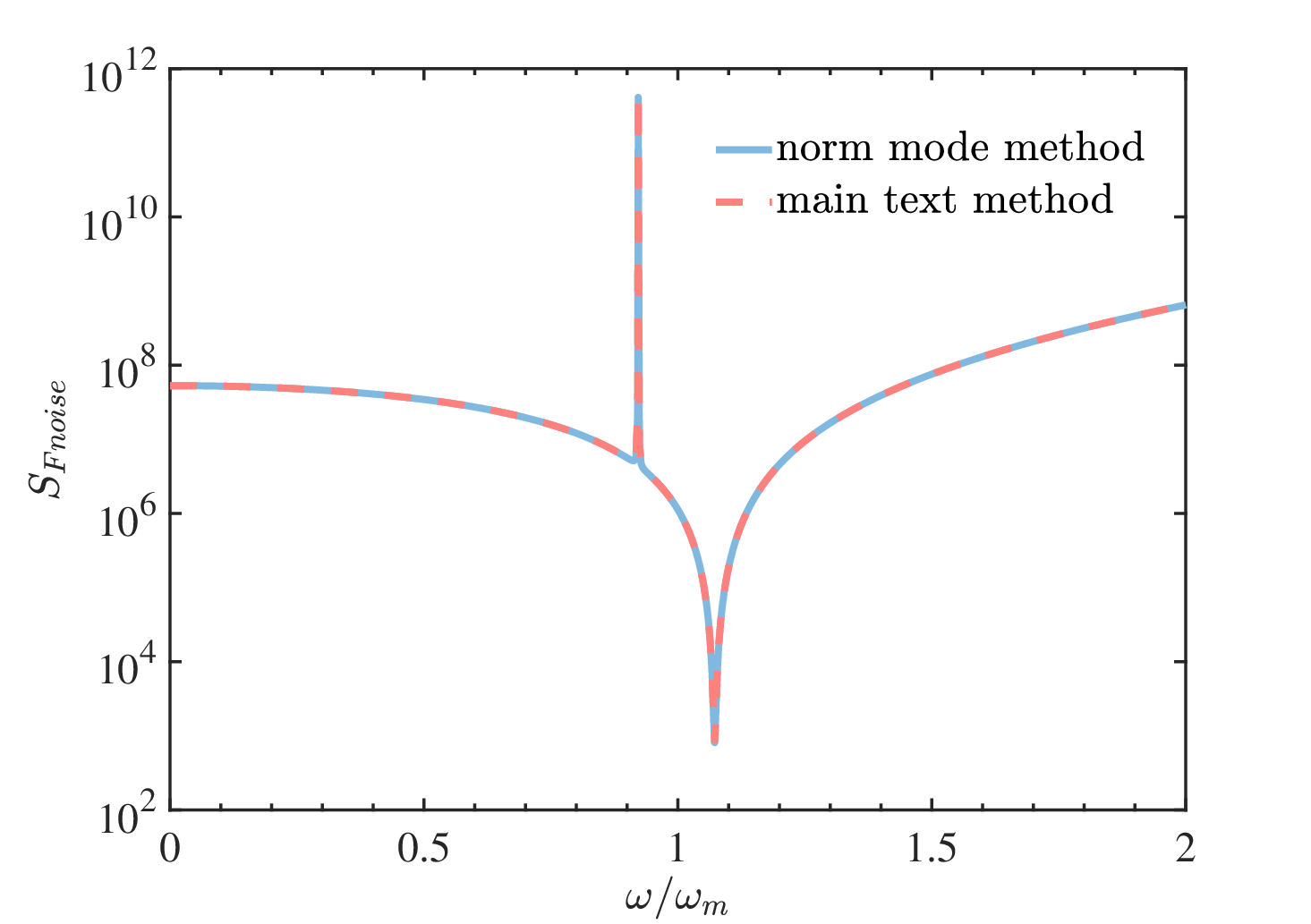}
	\caption{The noise spectral density $S_{Fnoise}$ with normalized frequency $\protect \omega /$ $\protect\omega _{m}$ based on the normal mode method and main text method.}
	\label{FIG10}
\end{figure}

We employ the definition of noise spectral density as provided in the main text and present the numerical simulation of this outcome. In Fig. \ref{FIG10}, It is obvious that both methods match well in terms of the position and size of the two norm modes.

\medskip
\textbf{Acknowledgements} \par 
This work was supported by the National Natural Science Foundation of China under
Grant No.12175029, No. 12011530014 and No.11775040.

\bibliographystyle{MSP} 
\bibliography{fsensing}  
\medskip


\end{document}